# Prediction of PEV Adoption with Agent-Based Parameterized Bass Network Diffusion Model

**Jarvis Yuan, Yihua Zhou, Jonah Lin, Kai Jin**

# Introduction

The adoption of vehicle electrification has been regarded as one of the most important strategies to address the rising concern in energy dependence and to mitigate the effect of climate change in the near future. In the last decade, although empirical data reviews are still suggesting that wide adoption of plugin electric vehicles (PEVs) are still difficult to achieve in short-terms, incentives from both the government and OEMs in the form of tax credit and additional packaged technology features are starting to attract and penetrate the early adopting population among the general public.

As a rising technology trend and innovation in its relative early stage, we hope to research and model the growth of a few specific regional PEV adoption, then apply and validate our models in a comparison study among different regions in the U.S.. In the process, we seek to gain a deeper understanding of related market-penetration factors and throughout the years in the form of an Agent-Based Bass Diffusion Model.

Our focus will be on the amount of influence that personal income has on an individual's decision to purchase an PEV. To achieve that goal we have obtained data from states that contain the PEV registration data for every electric vehicle purchased since 2010, as well as census data containing the income level and other various factors that are unique to every individual PEV adopter. The details of the dataset will be further discussed later.

To further facilitate PEV usage and reduce the lingering concern in range anxiety, public charging and supporting infrastructure need to be planned and built in advance, along with streamlining the production and reducing costs of components, among other factors. However, similar to the Chicken-or-Egg problem, OEMs, state agencies, as well as electric vehicle supply equipment (EVSE) suppliers wouldn't be able to increase production or incentivize purchase without a confident forecast of the adoption of PEV. Correctly predicting the number of electric vehicles, therefore, is essential in determining how much infrastructure we should build and where it should be built.

# Data and Methods

## Introduction of Agent-Based Model (ABM) and Bass Diffusion Model

The conventional approach toward modeling the spread of an event or in our case the adoption of PEV vehicles is often done with an numerical SI model, where the rate of the phenomena spreading is strictly controlled by a fixed variable β that controls the rate of spread which represents the probability of transmitting disease between a susceptible and an infectious individual. While this approach is useful in modeling a general spread of disease, it lacks the consideration for the circumstances of each individual in a spatio-temporal context. Our approach to modeling the data draws heavy inspiration from the combination of agent-based modeling (ABM) and Bass Diffusion Model, referenced in Figure 1. There are previous literatures that utilized the Bass Diffusion Process using an Agent-Based approach, where during each time step each individual has a unique variable that indicates chance to infect a nearby neighbor based on various factors. This is a more complex approach to the traditional use of an SI model as it integrates the additional dimensions in social interaction that simply do not exist in an SI model. This additional variability is more similar to what occurs in a real-world social network as each person has their own interaction circles, meaning every individual's chance of being infected should be different. This means at each time step the chance of an individual purchasing an PEV is different. The probability of infection then is determined by a mixture of inherent variables that are unique to the individual and also several global variables that affect everyone equally. For our diffusion model, the hypothesis is made that people have a higher tendency to adopt innovative technology including the PEVs when they discover that their friends or neighbors have adopted one, meaning that the rate of adoption should increase over time as more people decide to purchase a PEV. This concept builds twitter beyond that diffusion model applied in the Twitter adoption model used by Toole. Et al. [3], as in that experiment there are only two distinct chances of infection that is determined by whether an individual is an early adopter or not.

Prior to this project, there was a similar research conducted on the Electric Vehicles (EV) adoption in Morocco by Soumia et al. where researchers compared several innovation diffusion models including Gompertz, Logistics, and Bass, fixing EV sales growth with pricing of battery pack as the parameter for estimation [2]. The mentioned research seeked to find out which model would produce the most accurate forecast of EV adoption in the country, by comparing the predicted results to empirical data based on the R-square and Mean Absolute Percentage Error of each forecast model. A secondary objective was to assess whether the price of battery packs was significantly influential to the growth of the adoption. From the results the paper concluded that battery price was indeed a crucial determinant toward the adoption of EV's and that the Bass model produced the best fit when measured against empirical data. Their method of utilizing error metrics to measure the accuracy of the prediction will be also applied in this research to measure our modeled data against the empirical data.

```
Algorithm 1 Agent-based formulation of the Bass model
Require: number of agents M, external influence p, internal influence q, T
 1: x = (x_1, ..., x_M) ← (0, ..., 0)
 2: x̄ = (x̄_1, ..., x̄_M) ← (0, ..., 0)
 3: //Iterate over time
 4: for t = 1 → T do
 5:    //Iterate over agents
 6:    for all i = 1 → M do
 7:       p(adopt) ← β_{i,t}(k_1, k_2) = k_1 x_{1,i} + k_2 x_{2,i,t}
 8:       if rand = X(ω) ~ U(0,1) ≤ p(adopt) then
 9:          x̄_i ← 1
10:       end if
11:    end for
12:    x ← x̄
13:    adoptions_t ← Σ_{i=1,...,M} x_i
14: end for
15: return adoptions
```

Figure 1. Agent-Based Formulation of the Bass Model Algorithm [4]

## Data Gathering

As previously mentioned in the introduction, we're interested in the amount of influence that a person's income level has on that person's decision to purchase an PEV. To obtain that data, we needed records on the registration of PEV's over a time period, as well as a cluster of income data that can be matched to the individuals that made the purchase to the PEV. For the PEV registration data, we searched through the various state's DMV data to see if there's any database that contains granular data that contains records of individual PEV registrations. Among the various states in the United States that are considered early adopters of electric vehicles, Washington State was chosen as the empirical data source as the state has the most granular record vehicle registration data, listing out the vehicle type, VIN, registration date and zip code. The datatable shown in Figure 2 allows for the identification of every individual electric vehicle ever sold in the state of Washington since the beginning of the dataset, allowing for further exploration into trends of PEV adoption.

The dataset from Washington provides granular enough data that an agent based infection model can be established with a time increment in the weeks. From the dataset, a cumulative sum graph can be generated as empirical data of the total PEV registration over time per week in Figure 3, where the y-axis represents the absolute amount of cumulative PEV registration against the x-axis formatted by year-month-week_number.

| | Clean Alternative Fuel Vehicle Type | VIN (1-10) | Model Year | Make | Model | New or Used Vehicle | Sale Price | DOL Transaction Date | Transaction Type | Transaction Year | County | City | Postal Code | Electric Range | Base MSRP | Sale Date | State of Residence | 2020 Census Tract |
|---|---|---|---|---|---|---|---|---|---|---|---|---|---|---|---|---|---|---|
| 0 | Battery Electric Vehicle (BEV) | 5YJRE1A14A | 2010 | TESLA | Roadster | New | 0 | 2010-07-28 | Original Registration | 2010 | King | SEATTLE | 98112.0 | 245 | 110950 | June 25 2010 | WA | 5.303301e+10 |
| 1 | Battery Electric Vehicle (BEV) | 5YJRE1A14A | 2010 | TESLA | Roadster | New | 0 | 2010-03-17 | Original Registration | 2010 | Clark | VANCOUVER | 98664.0 | 245 | 110950 | February 01 2010 | WA | 5.301104e+10 |
| 2 | Battery Electric Vehicle (BEV) | 5YJRE1A14A | 2010 | TESLA | Roadster | New | 0 | 2010-12-22 | Original Registration | 2010 | King | REDMOND | 98052.0 | 245 | 110950 | December 11 2010 | WA | 5.303302e+10 |
| 3 | Battery Electric Vehicle (BEV) | 5YJRE1A17A | 2010 | TESLA | Roadster | New | 0 | 2010-12-22 | Original Registration | 2010 | King | MERCER ISLAND | 98040.0 | 245 | 110950 | November 28 2010 | WA | 5.303302e+10 |
| 4 | Battery Electric Vehicle (BEV) | 5YJRE1A18A | 2010 | TESLA | Roadster | New | 0 | 2010-06-22 | Original Registration | 2010 | King | KIRKLAND | 98033.0 | 245 | 110950 | May 31 2010 | WA | 5.303302e+10 |

Figure 2. WA PEV Raw Registration Data

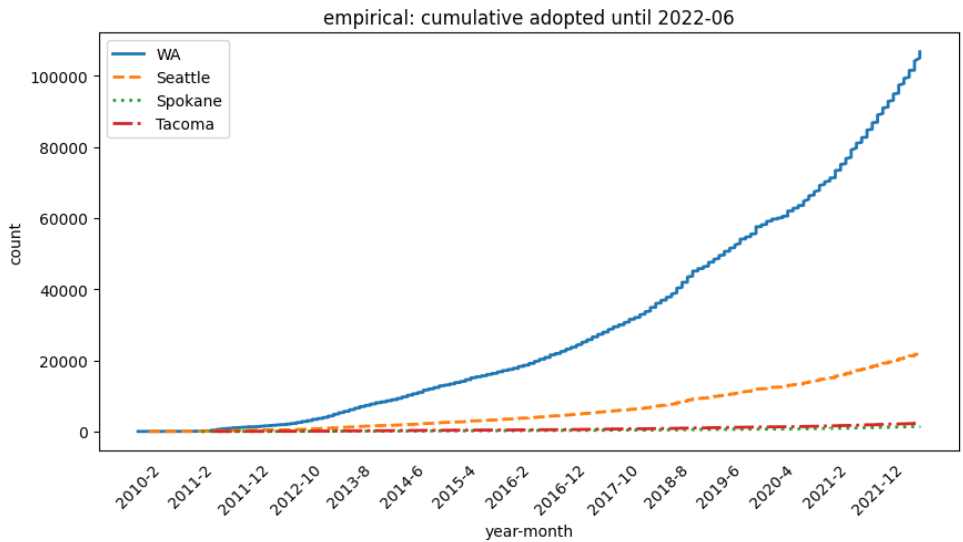

Figure 3. WA Weekly Cumulative PEV Registration (WA vs. Seattle vs. Spokane vs. Tacoma)

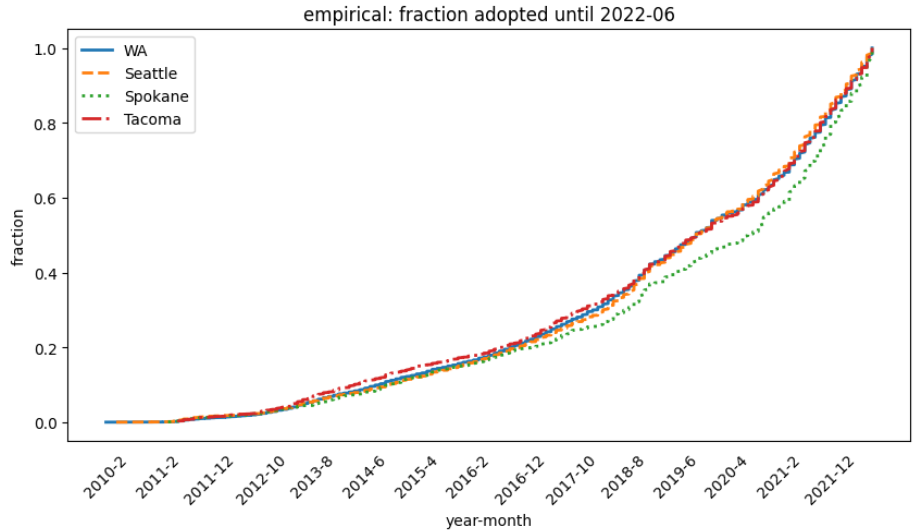

Figure 4. Comparison of Fraction Adopted (WA vs. Seattle vs. Spokane vs. Tacoma)

| Population | zip code | household count | <10,000 | 10,000~14,999 | 15,000~24,999 | 25,000~34,999 | 35,000~49,999 | 50,000~74,999 | 75,000~99,999 | 100,000~149,999 | 150,000~199,999 | >200,000 |
|---|---|---|---|---|---|---|---|---|---|---|---|---|
| 1209 | 99119 | 501 | 10.2 | 3.4 | 11.4 | 14.2 | 18.8 | 12 | 12.2 | 11 | 3.8 | 3.2 |
| 209 | 99128 | 82 | 3.7 | 8.5 | 14.6 | 4.9 | 7.3 | 28 | 9.8 | 20.7 | 1.2 | 1.2 |
| 5140 | 99141 | 2143 | 4.9 | 6.5 | 11.1 | 15.9 | 14.8 | 24.5 | 11.6 | 8.2 | 1.8 | 0.8 |
| 180 | 99149 | 88 | 11.4 | 13.6 | 17 | 14.8 | 28.4 | 9.1 | 0 | 4.5 | 1.1 | 0 |
| 390 | 99153 | 232 | 11.6 | 6 | 12.9 | 18.5 | 13.8 | 9.5 | 20.3 | 5.2 | 2.2 | 0 |
| 8122 | 99156 | 3416 | 10.5 | 7.9 | 12.6 | 13.5 | 13.1 | 20.1 | 12.2 | 6.1 | 2.9 | 1 |
| 938 | 99171 | 475 | 6.7 | 7.4 | 15.8 | 12.2 | 9.9 | 27.8 | 6.3 | 8.2 | 2.3 | 3.4 |
| 34174 | 98155 | 13261 | 4.1 | 3.2 | 8 | 5.3 | 9.2 | 18.3 | 15 | 19.8 | 8.9 | 8.1 |
| 538 | 99103 | 223 | 4 | 2.7 | 12.6 | 10.3 | 16.6 | 23.3 | 13 | 15.2 | 1.8 | 0.4 |
| 33822 | 99163 | 11300 | 19.2 | 9.3 | 15.2 | 10.8 | 10.2 | 10.3 | 9.5 | 8.6 | 3.7 | 3.3 |
| 3491 | 99343 | 780 | 0.1 | 4 | 14 | 15.5 | 5.8 | 26.7 | 23.8 | 6.8 | 0.8 | 2.6 |
| 4744 | 99003 | 1977 | 5.7 | 3.4 | 8.2 | 7.4 | 13.8 | 21.6 | 15.2 | 15.3 | 6.7 | 2.7 |
| 787 | 99012 | 333 | 3 | 8.1 | 9.6 | 10.5 | 19.2 | 18.9 | 9.9 | 14.1 | 6 | 0.6 |
| 13634 | 99016 | 5092 | 5 | 7.9 | 7 | 13.1 | 9.1 | 24.7 | 10.9 | 13.5 | 4.6 | 4.3 |
| 3174 | 99166 | 1375 | 12.9 | 9.7 | 13.5 | 6.5 | 16.7 | 14.5 | 14.1 | 9.2 | 2.8 | 0.1 |
| 183 | 99176 | 73 | 19.2 | 6.8 | 4.1 | 2.7 | 11 | 21.9 | 9.6 | 20.5 | 4.1 | 0 |
| 547 | 99179 | 215 | 1.9 | 0 | 6.5 | 8.4 | 8.8 | 23.3 | 31.6 | 11.6 | 7.9 | 0 |
| 12155 | 99006 | 4354 | 4.6 | 3.4 | 8.1 | 11.4 | 16.7 | 19 | 16.3 | 14.7 | 3.8 | 2.1 |
| 6169 | 99027 | 2390 | 6.4 | 5.3 | 10.6 | 10.3 | 13.2 | 16.9 | 13.3 | 16.4 | 5.6 | 1.9 |
| 731 | 99040 | 260 | 12.3 | 9.2 | 20 | 10.8 | 16.2 | 13.5 | 7.3 | 4.2 | 3.1 | 3.5 |
| 11046 | 98580 | 3974 | 7 | 2.2 | 10.3 | 9.3 | 9.4 | 25 | 11.3 | 15.1 | 7.3 | 3.1 |
| 36731 | 98584 | 13332 | 6.6 | 5.9 | 12.7 | 12.1 | 14.7 | 19.8 | 12.5 | 11.3 | 2.6 | 1.9 |

Figure 5. Population and Household Income Level Proportion per Zip Code

To generate zip code level socioeconomic values, we sourced income level and population data from ACS 2016 1-Year Estimates. Due to inflation and overall economic growth, we hypothesize that using constant household income data from either the start (2010) or end (2022) of the time horizon would misrepresent the actual purchasing power of the agents. We therefore chose the medium year (2016) where the census data is available. Combining the population and household income census data, the resulting data table in Figure 5, indexed by zip code in Washington, contains the corresponding zip code population and proportion of households in each income level. We have also broken down the data further by analyzing the cumulative PEV adoption growth of the 3 largest cities by population in the state, and here we see that Seattle's adoption curve is significantly higher than the other 2 cities.

For an individual to consider purchasing a PEV as their next personal vehicle, the purchasing power and the cost of PEV should both be considered. While household income level represents the purchasing power of agents on one side of the balance, the cost of PEV, in combination, would also account for the willingness of purchase on the other. Since the average price of a PEV varies across vehicle trims and OEMs (compact vs. luxury), we chose to represent the cost of PEV with unit battery pack price ($/kWh) as a global variable for our ABM. Further detail will be discussed in the "Future Work and Conclusion" section.

To support network link initialization, we sourced a lookup table containing the centroid coordinates per zip code. The rationale for using the centroid of each zip code will be explained later in the "Link Initialization" section under "Model Introduction".

## Model Introduction

| Variable | Definition |
|---|---|
| $\beta_{i,t}$ | Probability of adopting PEV for agent $i$ at time $t$ |
| $k_1$ | The adoption coefficient for income |
| $k_2$ | The adoption coefficient for number of neighbors who have adopted PEV |
| $x_{1,i}$ | Normalized value of income for agent $i$ |
| $x_{2,i,t}$ | Normalized value of the number of neighbors who have adopted PEV for agent i |
| $\hat{y}_t$ | Cumulative sum of total PEV adoption from model at time $t$ |
| $y_t$ | Cumulative sum of total PEV adoption from empirical data at time $t$ |

Table 1: Model Parameters and Definition

### (i) Node Initialization

The first step toward establishing a network is the initialization of nodes, which in this experiment represents a sample population in the state of Washington. For our experiment, we first identified the two types of variables which are global influential variables and individual influential variables. Individual variables are factors that are unique to the individual agent, while global variables affect every agent uniformly. Factors such as income, age, and average commute distance can all be considered individual variables while time-series values such as the pricing of gasoline and unit battery pack are considered global variables. For this project, we are specifically interested in the effect of household income on a person's decision to purchase an PEV.

We hypothesized that since electric vehicles are a relatively new form of transportation, the initial cost to adopt them would be high because of the scarcity of demand. However as more people purchase electric vehicles it will induce a demand for auto manufacturers to produce more and cheaper vehicles so they can sell their PEVs to more people. So to purchase an PEV there must be a level of influence from one's personal income level that strongly influences the person's decision to purchase an electric vehicle.

Our network is currently initialized with a population of N = 30,000 agents, representing around 0.5% of the total population size of Washington state. In comparison to the Twitter adoption model where the number of nodes is determined by the number of Twitter users in the final snapshot [3], our scaled population sample allows us to predict the future PEV cumulative adoption past our available registration data. Each agent is assigned to a zip code in WA in proportion to the empirical population per zip code. The larger population a zip code has, the more agents will be put under that zipcode. The equation for populating the nodes is therefore:

$$P_{zipcode,\ model} = P_{model} * \frac{P_{zip}}{P_{total}}$$

In terms of the initial state of adoption, we looked at the cumulative proportion of PEV purchases per zip code over the first year of the dataset, so a zip code that has more PEV registrations would be represented by a proportional amount of nodes that have the state of adoption. For example, if in the PEV registration dataset there exists 30 PEV in the beginning of time, zip code A has 10 PEV and zip code B has 20. Then, initially, the number of nodes that adopted PEV is consistent with that proportion, with the number of nodes in B twice the number of A. For our node initialization, we used the cumulative number of registered PEV after one year of the database establishment instead of the beginning of the dataset to avoid outliers and give the database a buffer to normalize to the future trend.

Each agent in the network has an income level ($x_1$) as an individual attribute. We hypothesize that income level, especially during the early and regular adoption periods of innovative products like PEVs, has a positive correlation with the probability of adopting a PEV itself. To further capture the effect of "word-of-mouth" in innovation diffusion and the adoption of PEVs, we added the second node attribute, number of adopted neighbors ($x_2$). While $x_1$ is assigned at the time of node initialization and remains constant for each node throughout the simulation, $x_2$ will be dynamically updated at each time step t, which will be further explained in the "Simulation Dynamic" section below.

(ii) Link Initialization

Individual agents then follow a probability-based rule to connect as links to form a social network. From Liben-Nowell et al. [1], the probability of a link being generated between two nodes can be modeled by a truncated power law function of distance δ (km):

$$P(distance\ \delta) = \delta^{-1.2} + v,\ where\ v = 5.0 * 10^{-6}$$

For our model, the distance δ between agent A and B is determined by the inter-zip code Great Circle distance (km) between the centroid coordinates of the zip codes assigned to A and B, respectively. For each zip code in the matrix, there exists (number of zip codes - 1) distances, see Figure 6 below.

| Zip Code | 98001 | 98002 | … | 99403 |
|---|---|---|---|---|
| **98001** | 0 | 2.663 | … | 248.349 |
| **98002** | 2.663 | 0 | … | 245.838 |
| **…** | … | … | 0 | … |
| **99403** | 248.349 | 245.838 | … | 0 |

Figure 6. Sample Inter-zip code Distance Matrix, *M*

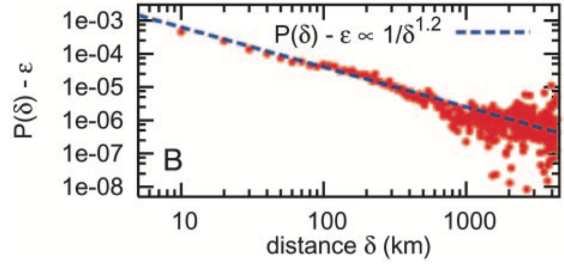

Figure 7. The relationship between friendship probability and geographic distance [1]

Intuitively, nodes that are closer to each other by distance have a higher probability of forming an acquaintance connection between them, and such probability decreases as the inter-agent distance δ increases until a background governing probability, $v$, takes over at large distances. It models the fact that in society the majority of connections between people come from their neighbors and friends who are closer in proximity of residence. Since the granularity of our agent network is at the zip-code level, we assumed that all agents assigned under the same zip code (δ = 0) would be fully connected in the social network.

(iii) Simulation Dynamic

Now that the network has been created, the simulation will be conducted in time steps of weeks, $t$, following the proposed algorithm in Figure 1 by Kiesling et al. [4]. At any given time, the adoption status of each agent is tracked by a binary state variable, susceptible (S) or infected (I), which corresponds to whether the agent has adopted an PEV. This contagion spreading is simulated by a mechanism resembling the Susceptible - Infected (SI) model, which is also a special case of the Bass model. In particular, we assumed an infected agent will remain infected until the simulation is reset, resembling the nature of purchasing a new vehicle as a relative long-term investment.

At $t = 0$, a small fraction of the agents are seeded to be initialized as infected. We follow the process of distributing the first infected agents based on empirical data, that is, zip codes with higher density of early adopting population in the registration data will have more agents initialized as infected. The probability of the susceptible individual being infected is determined by the variable β, a weighted linear combination of **personal income** ($x_1$) and **number of neighbors who have adopted and PEV** ($x_2$) at a specific time-step, formulated as:

$$\beta_{i,t}(k_1, k_2) = k_1 x_{1,i} + k_2 x_{2,i,t}$$

, where *i* denotes each agent and *t* represents time step during the simulation, and coefficients $k_1$ and $k_2$ determine the level of influence that each individual variable has on the overall probability of infection for the individual. At each time step, each susceptible agent's adoption status is evaluated and a new $\beta_{i,t}$ is assigned based the new value of $x_2$, as long as the individual is still susceptible (S). This linear formulation of the adoption probability has several advantages; It first allows flexibility in appending new adoption variables, average commute distance (m), for example, and allows easy fine-tuning of the corresponding coefficients, to capture more dynamic in the diffusion process. Secondly, the model is no longer confined to relying on processing of empirical adoption data to assign "early adopters" or "regular adopters". It models the adoption probability uniquely to each agent in the network relative to their personal and social attributes. Lastly, since census data is readily available from API connections and the number of connected neighbors is inherent to the social network, this formulation would also allow for easier scalability and transferability when considering expanding the study to other states or regions of interest.

To validate the model diffusion of PEV in the state of Washington, we will step through the model for T=612 week steps as a complete simulation, which represents the number of weeks in our available PEV registration data after initialization. Again, compared to the methods of parameter estimation of the diffusion model equations [2], the combination of ABM and the SI model would not only allow us to obtain the predicted cumulative and weekly new PEV registration, but also allow us to peek into the progression of PEV adoption among agents with different socioeconomic and social network characteristics.

Given our relatively large network degree and rule-based link connectivity, we decided to use the open-source python library, [NetworKit](), to establish our social network. From [an online benchmark report](), NetworKit can out-perform NetworkX by at least 20X in runtime using a set of standardized computation exercises.

## Parameter Estimation

Once we have obtained simulation results, we can now regressively evaluate the accuracy of the result compared to empirical data using the error rate metric of choice, Mean Absolute Error (MAE). For each time step in the simulation, the MAE of our prediction is obtained by comparing the difference between the cumulative sum of the empirical and predicted adoption data.

$$\text{MAE} = \frac{abs|y_t - \hat{y}_t|}{T}$$

For the previous paper done on the Morocco EV adoption, they used least squares to numerically estimate the global diffusion parameters of the generalized bass model [2]. While a least-squares approach to fixing the global diffusion parameter would be ideal for our parameters as well, due to our agent-based modeling approach it presents an inherent nonlinearity during the transition steps and therefore a least-squares method would not apply. Specifically, our model's adoption coefficient $k_2$ is influenced by a constantly changing value of the number of infected neighbors, making it non-linear over each time step. Therefore we devised some reasonable constraints and utilized the method of random search to reduce the complexity of this search process for the optimal model adoption coefficients.

Given our weighted linear combination of β, the adoption variables need to be normalized before fitting the adoption coefficients. For the attribute of income, we divide the income of each agent by the value of the largest income group, which is about 200k. In terms of the number of neighbors who have adopted electric vehicles, we divide that value by the largest degree in the graph, knowing that the number of neighbors will not exceed the number of degrees [k], thus guaranteeing this value is also in the range of [0-1]. We know that the upper bound of $β_{i,t}$ would be constrained between 0 and 1. The generalized function to normalize income and number of adopted neighbors is:

$$x_{i,norm} = \frac{x_i - x_{max}}{x_{max} - x_{min}}$$

Beyond confining $0 < β_{i,t} <= 1$, we can also add constraints to $k_1$ and $k_2$ through empirical reasoning, and further reduce the complexity of the estimation. Here we can assume that the search space for $k_1$ and $k_2$ are the same since both come from the same distribution of values, i.e. $k_1, k_2 \in \{0, 0.1\}$ with fixed cardinality. If we search through 100 values for each $k$ then the number of permutations is calculated by $\frac{100!}{(100-2)!} = 9900$, which is significantly less than searching through a model with no constraints. At each time step, the model's fitness is determined by comparing the predicted cumulative sum of the number of adopted neighbors to the empirical number using the MAE metric MAE = $\frac{abs|y_t - \hat{y}_t|}{T}$ where $\hat{y}_{t'}$ is the predicted sum and $y_{t'}$ is the empirical sum.

In terms of searching the best combination of parameters, we use the method of random search. Firstly, we do a manual naive search to get a reasonable estimate of parameter scope. And then iteratively test each combination with random. If we can find a better pair of parameters, then we manually shrink the scope and do random search again. Once we can not find a better combination of parameters with 100 iterations, then we will stop and regard that we found the best pair of $k_1$ and $k_2$ that fits the model.

# Results

Cumulative Adoption Evaluation

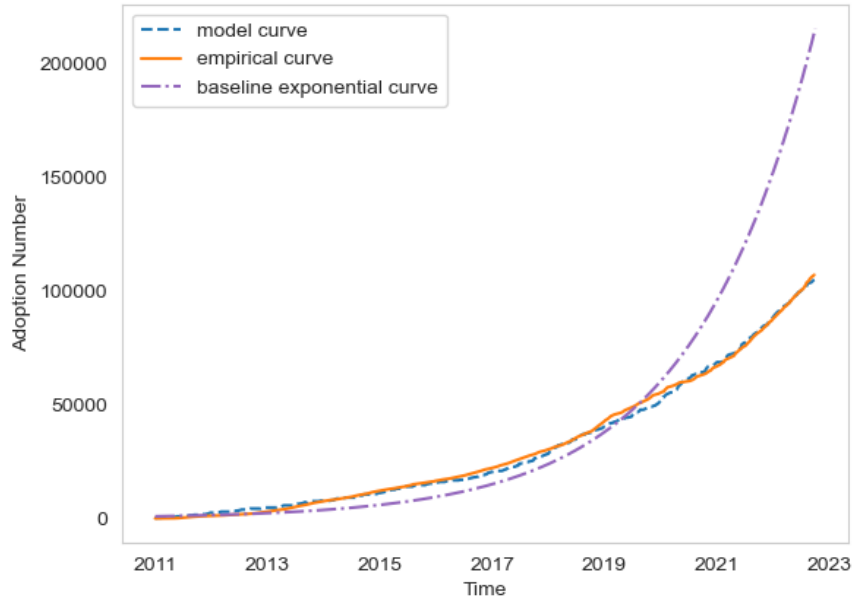

Figure 8. Predicted model vs. Empirical data vs. Baseline Exponential Fit (30k agents)

Our ABM network consists of 30k agents connected by 6.9 million links, which is roughly 1.4% of all possible link combinations between nodes after the link initialization filtering. While the network initialization took around 15 minutes from scratch, using a saved edge list, our NetworKit undirected network can be constructed in less than 5 seconds.

For the x-axis of Figure 8, one step (t) represents one week, with the first step representing the first week of 2011 and the last step (612$^{th}$) representing the week in August 2022. For the y-axis, the value is the number of total PEV adoption in the state of Washington. Figure 8 compares our simulation model and an exponential fit model against the empirical data. Among the models, the exponential baseline model fitted the worst. This model growed slower than the empirical curve in the early stage, but the constant growth acceleration led to much faster growth later on. The result of this exponential model ended up almost twice as much as the empirical data in the last step.

After running the simulation at each time step, using random search we were able to find the optimal parameters for $k_1 = 9.28 * 10^{-6}$ and $k_1 = 8.43 * 10^{-3}$ under 30k agents which results in best fits seen in Figure 8. Using these optimal values for $k_1$ and $k_2$, we were able to achieve an MAE of around 1000 (unit: count of PEV registration) for the last time step at t=612, roughly **1%** of error. Also, On average the absolute error per step is about 1500.

Figure 8 shows great consistency between the empirical data and the simulated model data. The time period from 2020 to mid 2021 has a relatively large difference between these two curves of data, which is reasonable as we suspect that this is due to the influence of the COVID-19 pandemic and in our assumption we did not take that into consideration. Except for this period, our model is able to closely and consistently resemble the empirical data in terms of cumulative sum of registered PEV, demonstrating that the income and the number of neighbors adopting PEV vehicles have a significant effect on the diffusion of PEV adoption.

We theorized that in the initial phase of adoption, agents who are in the **higher income bracket**, therefore, have high purchasing power, and those with a **high degree of social connection,** therefore, a large degree of social influence, would be more inclined to purchase in PEV. Through this simulation, we have demonstrated that early adopters are more likely to come from zipcodes closer to big cities because the income level is reasonably higher and there are more social connections in the cities to jointly push the diffusion of new innovation among residents. This is somewhat similar to the results found from the twitter adoption experiment from Toole et. al [3] where they have discovered that critical mass achievement, the measurement of the shift from early adopters to mass adoption often happens first in larger cities. Our results have shown similar findings where the wave of early adopters are often located near big cities, which could be caused by their better access to PEV charging infrastructure and also higher overall income.

## Critical Parameter Analysis

We also conducted an investigation into what the fit of the model would be if we only used the individual parameters of income and number of adopted neighbors to fit the data. Running the same procedure of parameter estimation as previously, we simply set the other variable's coefficient to be 0 to refit and tune the model using a random search scheme until the termination condition is met. In Figure 9A, we can see that using income data alone, the shape of the curve does not resemble the growth curve of the empirical model, and no matter what coefficient we use for income it will never be the same shape as the empirical model. Similarly for Figure 9B, the number of adopted neighbors model under-predicts the model greatly, and does not resemble the shape of the empirical model. In contrast, using both the income and number of adopted neighbors as parameters created a model that closely resembles the empirical model in Figure 9C. The comparison of results is displayed in Figure 9D. What we can see here is that using the income model alone over predicts the model to a certain extent, and using the number of adopted neighbors model alone under predicts the model extent. We theorize that the effect of income therefore plays a stronger influence on an individual's decision to purchase an PEV, but after the adoption has reached a certain timestep, perhaps the saturation for critical mass has been achieved, the "word-of-mouth" effect as the result of number of adopted neighbors model becomes stronger in determining if a person is going to purchase an PEV or not. In Table 2, the numerical values of MAE are shown and using a combination of the two variables gives significantly lower MAE.

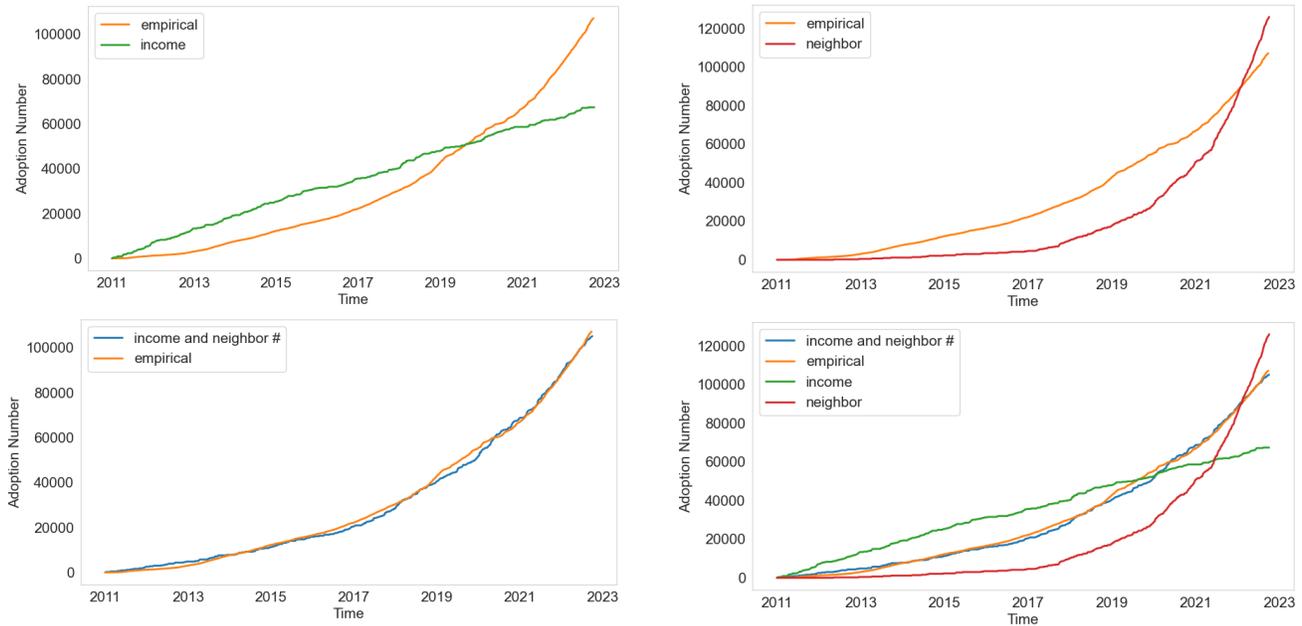

Figure 9 (A: Top Left, B: Top Right, C: Bottom Left, D: Bottom Right). Aggregated Model Types vs. Empirical Data (30k agents)

| Model | Baseline | Income | Neighbor # | Income and Neighbor # |
|---|---|---|---|---|
| **MAE** | 39,073 | 10,902 | 12,661 | **1,327** |

Table 2. Aggregated model types vs. Empirical data MAE Results (30k agents)

This reveals a critical issue in the adoption of PEV for the masses as many suburban dwellers will tend to not choose PEVs if there is not an abundance of charging stations in their vicinity, which will lead to an adoption gap between the two types of populations. This could be alleviated by the installation of home based PEV charging stations. Another issue presented then is the gap between the rate that people are purchasing PEVs and the rate that PEV charging stations are getting installed in the cities. If cities are the source of the majority of growth in PEV adoption, then cities must invest an equal amount of funding in building PEV charging stations. As of now there are not many apartments that offer PEV charging stations in the complex, and drivers often have to travel a certain distance to access charging stations. This could potentially disincentivize a potential buyer into purchasing an PEV.

## Spatio-Temporal Distribution Analysis

As mentioned in the section "Introduction of Agent-Based Model (ABM) and Bass Diffusion Model", one of the benefits of combining ABM with a diffusion model is the ability to capture the spatio-temporal dimension in the diffusion process. In Figure 11, we visualized the PEV adoption heatmap comparing the empirical and model prediction at the final time step (T=612) of the available empirical PEV registration data. In the heatmap, each dot represents an agent aggregated onto its assigned zip-code spatially, where a high concentration of the adoption region is represented by the density of the clustered dots. Playing back in time, we observed that our model is able to closely capture both the density and distribution of new PEV adoptions when compared to the time-series heatmap from the empirical data, and even the low to low density areas from the empirical heatmap looks to be closely resembled. From the time-series heatmap, we could also notice that PEV adoption tends to be concentrated in the larger urban cities, such as Seattle, Olympia, Portland (bordering with Oregon), and Spokane, and its surrounding suburban areas, which we will further explore in the next section.

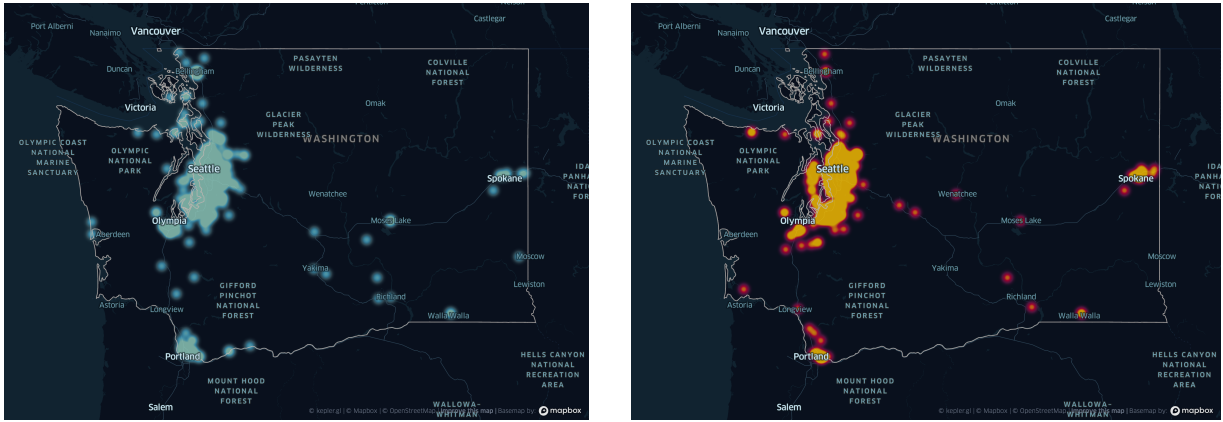

Figure 11. Empirical (left) v.s. Model Simulated (right) Adoption Heatmap at T=612

## Critical Mass Achievement Evaluation

Inspired by Twitter Adoption Paper by Toole et. al [3], we will also visualize locations of zip codes which achieved critical mass with two series of temporal snapshots as our next step. Critical mass is a key moment for any contagion process; Similar to the Twitter Adoption Paper, we mark a county as reaching its critical mass when 13.5% of all eventual PEV registrations took place, indicated from the empirical data. The first series will be from empirical data, and the second series will be from our simulation result. So, the geographical adoption spread of our model can be compared to the empirical data. In the Twitter Adoption Paper, a dashed line of x=y means a perfect match between the empirical and simulation data. If a county's prediction is to land on that line, it means the predicted duration to achieve critical mass is the same as the empirical duration to achieve critical mass. Together in Figure 12, Figure 13, and Figure 14, we can see that the populous counties such as Pierce, King, and Snohomish county are all closer to the dashed line, and we believe that it is because these counties have larger agent volume and are more densely populated which allows them to amplify the effect of "word of mouth" and therefore better resemble the empirical growth. As for reasons on why some counties are very far from the line, we discovered that these counties inherently have a low population density. Due to the low resolution of our model, one agent adoption during simulation imitates 200 registered PEV by population in the real world. Though our model could well-predict the time-series cumulative adoption as an aggregate of the entire state, this discreteness of the simulation means that the accuracy would break down when we zoom into smaller aggregate units such as counties or zip-codes. Nevertheless, we can conclude that larger counties with a higher population density are more likely to achieve critical mass first.

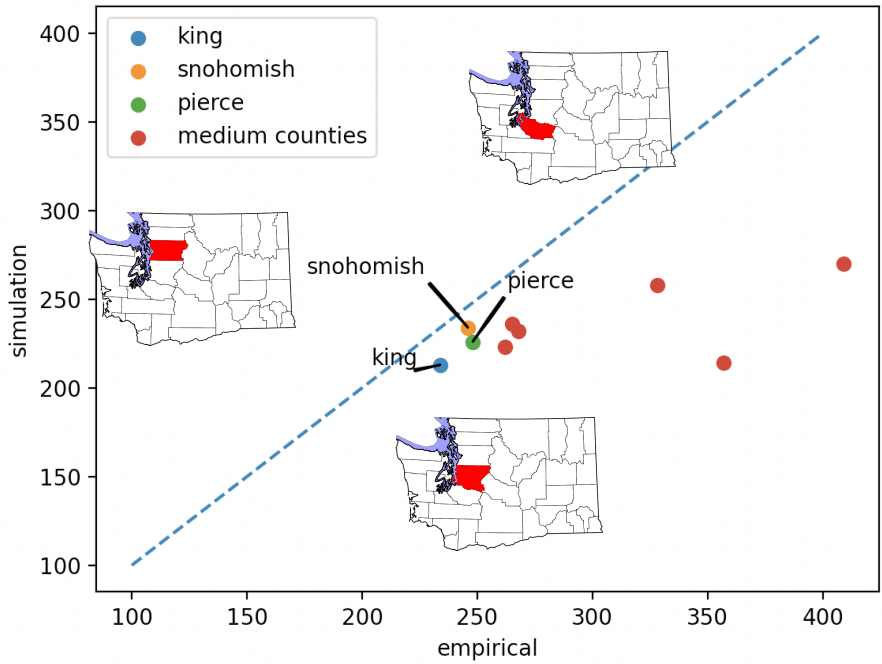

Figure 12. Week to reach C.M. by Top 10 counties in PEV registration in WA: Empirical vs. Simulation (30k agents)

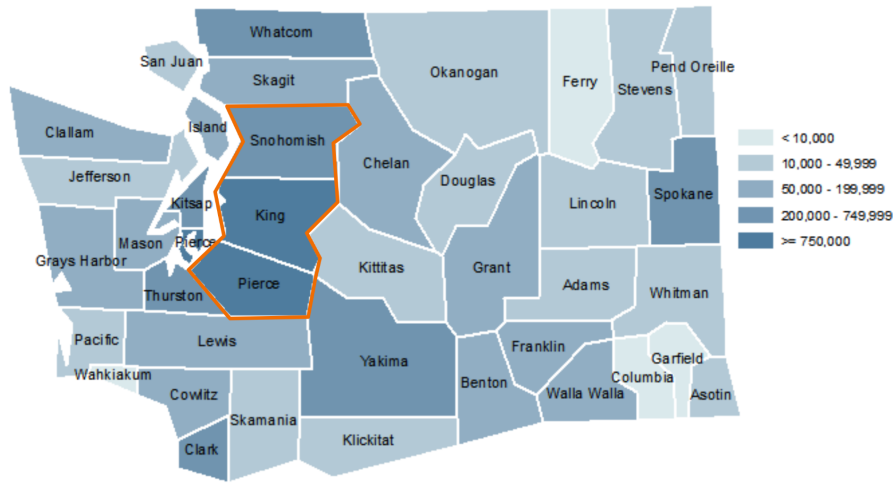

Figure 13. Population Distribution by County in Washington

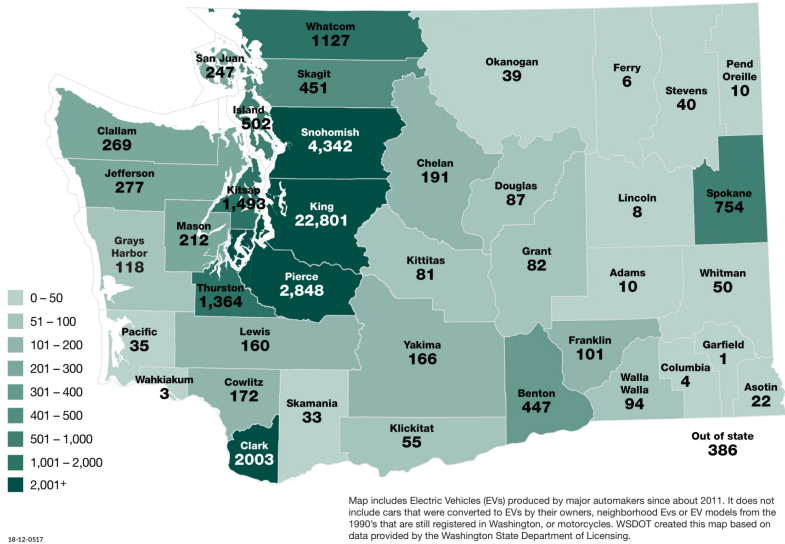

Figure 14. "Washington State EV Registrations Top 40,000"
http://www.pluginncw.com/news/2019/1/23/washington-state-ev-registrations-top-40000

## PEV Penetration Prediction

With the model being able to predict the empirical growth to an acceptable level given our available PEV registration data until June-2022, we also ran the simulation beyond the empirical data's timeline to see at which point the population would reach a 5% PEV penetration.

$$\text{PEV penetration \%} = \frac{\text{Number of PEV}}{\text{WA Population}} * 100\%$$

In Figure 15, we have discovered that with the current growth trends in adoption, it is likely that the adoption is still in the early stage so the growth will grow at a faster rate over time before reaching its linear growth and plateau stage. We predicted that the 5% PEV penetration will be reached in June-2027.

Referencing Senate Bill 5974, 2022, "All light-duty vehicles sold, purchased, or registered in Washington state must be EVs by model year 2030. The Interagency EV Coordinating Council must develop a plan for achieving this goal by December 31, 2022." According to our model prediction, the PEV penetration would likely be below 10% by year 2030.

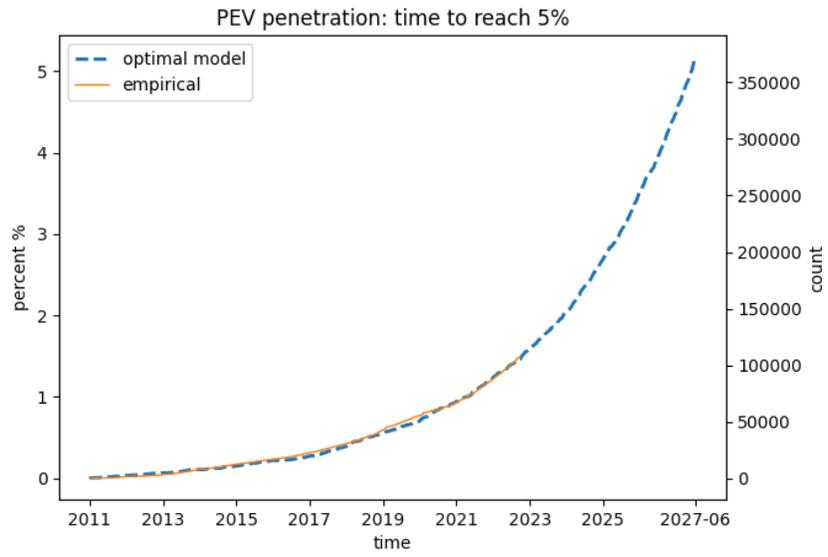

Figure 15. Prediction of PEV Penetration to Reach 5%

# Future Work and Conclusions

From our studies, we learned a few critical points about the adoption behavior of PEV owners over time in both a spatial and temporal trend. In terms of temporal trends, we found that personal income is a strong factor in the probability of adopting a PEV but only during the early stages of adoption, once adoption has reached a certain point the influence of nearest neighbor becomes stronger than personal income. This reveals a critical issue in the future adoption of PEVs, as the income gap will certainly be an obstacle in the early adoption stage of PEVs, and therefore governments should focus on prescribing policies to increase accessibility to PEVs for the lower income demographic.

Using the temporal analysis of our simulation we also found that the adoption occurs at a much faster rate in urban areas where population density is high, which means the infrastructure will inevitably become the bottleneck for growth in urban areas. We believe that legislation should address accessibility in urban centers first, since these areas will see the greatest growth in demand for electric vehicles. This means stronger incentives for building PEV infrastructure needed to support the growing network, as well as clean energy rebates to encourage clean energy charging.

Our simulation focuses on utilizing the network model to simulate the growth in PEV adoption, and can be used to predict beyond empirical data into future demands in PEV adoption. For future work, this approach can be compared to other methods of prediction such as line fitting for the evaluation of accuracy. The model's accuracy should also be tested against future empirical data once those have become available. If the prediction can be verified to be accurate, then it can be applied to model and project the PEV adoption rates of other states outside of Washington, such as California, another early PEV adopting state with more progressive policies towards transportation electrification.

From the simulation, we found that personal income has a strong influence on the probability of purchasing an PEV, however it is certainly not the only factor. We have also theorized that global variables such as lithium-ion battery pack pricing may also influence the adoption of PEV long term. While sufficient personal income is necessary to purchase a PEV due to their higher average market price compared to their gasoline counterpart, currently, the pricing of PEVs will soon decrease to a point that they'll be on par with ICE vehicles in price, especially taking into account of the lower maintenance costs due to the electric powertrain. As we can see from the empirical data in WA, this trend in PEV adoption is certainly still in the early phase, as Washington State is one of the earliest adopters of electric vehicles in the country. To support such a growth in PEVs on the road, infrastructure also needs to grow at the same speed which is a lot more difficult than simply buying an PEV. Building a PEV charger requires funds and space, and in many urban centers there is often no space for apartment dwellers to access PEV charging stations. So beyond predicting the future trends of PEV growth, it is also important to utilize that knowledge to properly allocate resources to build out the infrastructure needed to support that growth. Otherwise, the adoption of PEVs will remain socially divided where only people who can afford a PEV and have access to chargers will consider buying a PEV.

# Appendix

A-1: GitHub Repository Link:
https://github.com/RichZhou1999/CE263-EV-Adoption-Diffusion-Model-

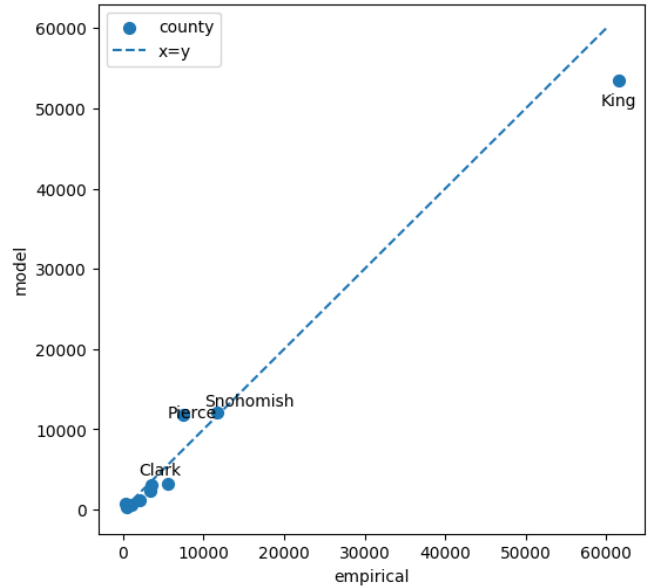

Fig A-2: Model vs. Empirical: Cumulative Adoption of Top 10 WA Counties to Reach Critical Mass

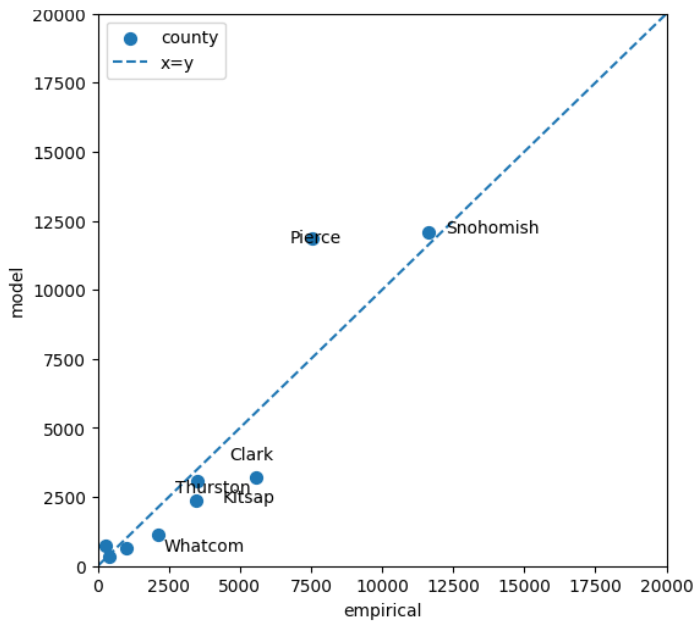

Fig A-3: Model vs. Empirical: Cumulative Adoption of Top 9 WA Counties to Reach Critical Mass, Zoomed-in (x<20,000, y<20,000)

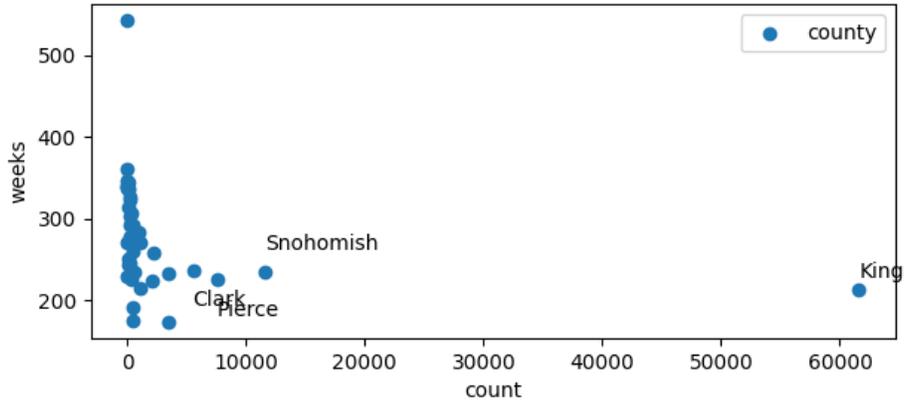

Fig A-4: Empirical: Total PEV Registration Counts vs. Week to Critical Mass

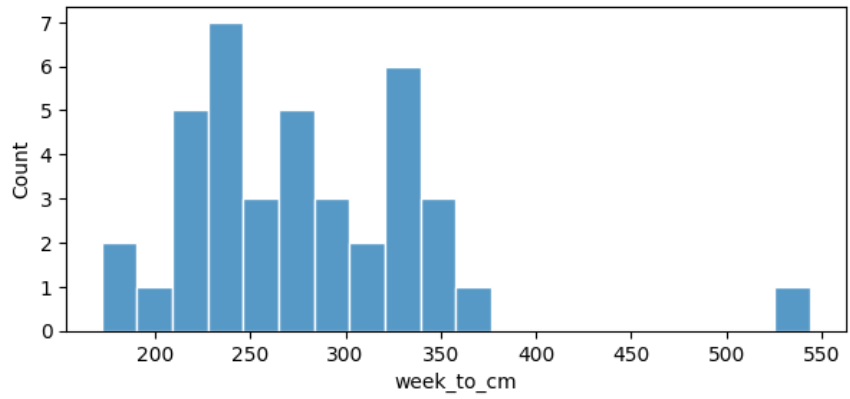

Fig A-5: Empirical: Histogram of Week to Critical Mass for All WA Counties